\documentclass[journal=tches,final]{iacrtrans}

\usepackage{booktabs}

\author{Corbin Hibler \and Firas Hassan \and Eric McKanna}
\institute{
  Ohio Northern University, Ada, USA\\
  \email{c-hibler@onu.edu} \and
  \email{f-hassan@onu.edu} \and
  \email{ewmckanna@gmail.com}
}

\title[BipBipCache]{BipBipCache: Pipeline-Aware Integration of Low-Latency Tweakable Encryption in an Embedded Cache Controller}

\begin{document}

\maketitle


\begin{abstract}
Consumer and embedded processors store sensitive data in on-chip SRAM caches that remain readable after power loss or physical probing unless ciphertext is maintained in the memory array itself. This paper presents BipBipCache, a direct-mapped cache controller that integrates the BipBip tweakable block cipher (TBC) to encrypt cache data and tags in real time using a C$^3$-style 24+40 bit decomposition of each 64-bit word. We reconstruct the first pipelined hardware BipBip encryptor from a decryptor-centric specification and coordinate it with a 3-cycle decryptor inside the cache datapath.

Our threat model targets confidentiality of cache-resident contents against cold-boot, bus, and SRAM readout attacks---not microarchitectural side channels, which require randomized caches rather than encrypted ones. A key architectural result is that 6-cycle encryption latency does not fully translate into 6-cycle write penalty: the first three encryptor stages overlap with tag decryption and hit detection, leaving an effective 3-cycle write commitment after hit verification. We verify encryptor and decryptor correctness against the official BipBip C++ reference (five vectors each), report FPGA resource utilization on Xilinx Artix-7 (3\,356 LUTs, 16.1\% of device; crypto logic $\approx$79\% of LUTs), and confirm end-to-end operation on hardware.
\end{abstract}

\section{Introduction}
\label{sec:introduction}

Low-cost embedded and consumer devices---IoT SoCs, smart peripherals, and portable electronics---increasingly process credentials, personal data, and firmware secrets in on-chip SRAM. Unlike main memory, cache contents are rarely encrypted in commercial microarchitectures, leaving them exposed to cold-boot attacks~\cite{haldermanLestWeRemember2009}, direct SRAM probing, and offline observation of memory buses or modules. Hardware countermeasures that operate transparently at cache access latency are attractive for these platforms because software-only protections often impose unacceptable overhead.

A distinct line of work addresses \emph{microarchitectural} side channels such as Prime+Probe and Flush+Reload through \emph{cache randomization}: ScatterCache~\cite{wernerScatterCacheMitigatingCache2019} and SCARF~\cite{canaleSCARFLowLatencyBlock2023} encrypt or permute set indices so attackers cannot build reliable eviction sets. BipBipCache pursues a complementary goal: \emph{encrypted-cache} confidentiality, in which stored data and tags are ciphertext rather than plaintext. This protects against physical and bus-level readout but does \emph{not} mitigate contention-based side channels; we state that limitation explicitly in Section~\ref{sec:threatmodel}.

Several low-latency ciphers target memory and pointer protection. QARMA~\cite{avanziQARMABlockCipher2017} and MANTIS~\cite{beierleSKINNYFamilyBlock2016} were designed for memory encryption with competitive ASIC latency. PRINCE~\cite{borghoffPRINCELowLatencyBlock2012} optimizes single-cycle encryption in hardware. Cryptographic Capability Computing (C$^3$)~\cite{lemayCryptographicCapabilityComputing2021} introduced pointer encryption using ultra-low-latency primitives; BipBip~\cite{TCHES:BDDGR23} was designed for C$^3$ with a 24-bit block, 40-bit tweak, and 3-cycle ASIC decryption latency.

What distinguishes our work is the combination of (i)~the first pipelined BipBip \emph{encryptor} in hardware, reverse-engineered from a decryptor-only publication, and (ii)~a pipeline-aware cache controller that exploits asymmetric 6-cycle encryption and 3-cycle decryption to hide part of the write latency behind tag verification.

Our contributions are:
\begin{enumerate}
    \item \textbf{First hardware BipBip encryptor.} We invert decryptor round functions and realize a 6-cycle pipelined encryptor in VHDL, verified against the reference implementation~\cite{bipbipReferenceCode}.
    \item \textbf{Encrypted cache integration prototype.} We present a direct-mapped cache controller with three BipBip instances (data encrypt, data decrypt, tag decrypt) and document the 64-bit-to-(24+40) cryptographic mapping used on both data and tag paths.
    \item \textbf{Pipeline overlap analysis.} We show how tag-path decryption synchronizes with the encryptor so that only three additional cycles contribute to effective write latency after hit detection.
    \item \textbf{FPGA evaluation.} We report synthesis and implementation results on Artix-7, hierarchical crypto overhead, and board-level functional tests.
\end{enumerate}

\section{Background}
\label{sec:background}

\subsection{Tweakable Block Ciphers}

A tweakable block cipher (TBC) is a family of permutations $\widetilde{E}: \mathcal{K} \times \mathcal{T} \times \{0,1\}^n \rightarrow \{0,1\}^n$, where $\mathcal{K}$ is the key space, $\mathcal{T}$ the tweak space, and $n$ the block size~\cite{JC:LisRivWag11}. For fixed key $K$ and tweak $T$, $\widetilde{E}_K^T(\cdot)$ is a permutation on $n$ bits. The public tweak diversifies encryptions without changing the secret key---natural for memory encryption when each address or context supplies a distinct tweak~\cite{rogawayEfficientInstantiationsTweakable2004,chakrabortyGeneralConstructionTweakable2008}.

\subsection{The BipBip Cipher}
\label{sec:bipbip}

\begin{figure}[ht]
    \centerline{\includegraphics[width=0.85\textwidth]{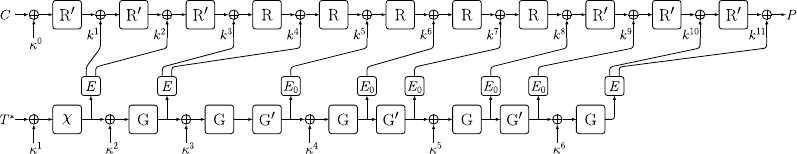}}
    \caption{BipBip high-level decryptor structure (adapted from Belkheyar et al.~\cite{TCHES:BDDGR23}).}
    \label{fig:bipbip-structure}
\end{figure}

BipBip~\cite{TCHES:BDDGR23} is a TBC with $n{=}24$, $|\mathcal{T}|{=}40$, and a 256-bit master key, designed for C$^3$ pointer encryption. Figure~\ref{fig:bipbip-structure} summarizes its decryptor-oriented structure; we refer to Belkheyar et al.~\cite{TCHES:BDDGR23} for round definitions ($R$, $R'$, $G$, $G'$) and key scheduling.

Our encryptor inverts the published decryptor datapath and runs the tweak schedule forward. Wang et al.~\cite{wangCryptanalysisFullRoundBipBip2024} later refuted the designers' 96-bit security claim with a full-round attack at $\approx 2^{92}$ complexity that remains impractical. We use BipBip as a case-study primitive; the integration principles transfer to other low-latency TBCs.

\section{Threat Model and Security Goals}
\label{sec:threatmodel}

BipBipCache is an \emph{encrypted cache}, not a \emph{randomized cache}. Table~\ref{tab:threatmodel} states our goals and non-goals.

\begin{table}[ht]
\centering
\caption{Threat model for BipBipCache.}
\label{tab:threatmodel}
\small
\begin{tabular}{p{0.46\linewidth}p{0.46\linewidth}}
\toprule
\textbf{In scope} & \textbf{Out of scope} \\
\midrule
Confidentiality of data/tags held in cache SRAM & Prime+Probe, Flush+Reload, Spectre-class leakage \\
Protection against cold-boot and physical SRAM readout & Set-index privacy (index bits not encrypted) \\
Binding stored words to tweak context via TBC & Strong authentication (no MAC; encrypted tag $\neq$ AE) \\
Hit requires decrypt-and-compare of stored tag & Active integrity against key-bearing adversary \\
\bottomrule
\end{tabular}
\end{table}

\paragraph{Why encrypt tags?} The logical address tag supplied by the processor is not stored verbatim. A hit requires decrypting the stored tag and comparing it to the (pipeline-delayed) incoming tag. An adversary who patches SRAM with arbitrary ciphertext cannot synthesize a valid hit without knowing the key and correct tweak binding. This is tamper-evident lookup metadata, not a cryptographic MAC.

\paragraph{Relation to randomized caches.} ScatterCache~\cite{wernerScatterCacheMitigatingCache2019} and SCARF~\cite{canaleSCARFLowLatencyBlock2023} target side-channel resistance by randomizing set mapping with low-latency ciphers on index bits. BipBipCache instead ciphertext-protects line contents and tags; the 7-bit set index remains in cleartext for direct-mapped lookup.

\section{System Design}
\label{sec:design}

\subsection{Cache Organization}

BipBipCache is a direct-mapped cache with a 64-bit address space decomposed as: 52-bit tag (bits 63--12), 7-bit set index (11--5), 2-bit word offset (4--3), and 3-bit byte offset (2--0). Each of 128 sets stores four 64-bit words (256-bit line), a 52-bit tag, a valid bit, and a dirty bit. Table~\ref{tab:metadata} summarizes metadata roles.

\begin{table}[ht]
\centering
\caption{Cache metadata and encryption coverage.}
\label{tab:metadata}
\small
\begin{tabular}{lcccl}
\toprule
\textbf{Field} & \textbf{Width} & \textbf{Encrypted?} & \textbf{Writer} & \textbf{Role in pipeline} \\
\midrule
Set index & 7 & No & Address & SRAM row select \\
Data word & 64 & Partial$^\dagger$ & Controller & Payload; 24 bits via BipBip \\
Tag & 52 & Partial$^\dagger$ & OS / ROM model & Hit detect after decrypt \\
Valid & 1 & No & OS / ROM model & Gates hit signal \\
Dirty & 1 & No & Controller & Write-back hint \\
\bottomrule
\multicolumn{5}{l}{\footnotesize $^\dagger$See Section~\ref{sec:cryptomapping}: 24 bits encrypted, 40 bits as tweak context.}
\end{tabular}
\end{table}

We use direct mapping to minimize control complexity while validating the cryptographic datapath. Tag and valid arrays are modeled as OS-provisioned ROM-like structures in simulation; the controller reads but does not rewrite them.

\subsection{Cryptographic Mapping}
\label{sec:cryptomapping}

BipBip operates on 24-bit blocks with 40-bit tweaks---not on full 64-bit words. Following the C$^3$ pointer layout~\cite{lemayCryptographicCapabilityComputing2021,TCHES:BDDGR23}, each 64-bit cache word $W$ is decomposed as:
\begin{align*}
  T &\leftarrow W[63{:}58] \mathbin\| W[33{:}0] && \text{(40-bit tweak)} \\
  P &\leftarrow W[57{:}34] && \text{(24-bit plaintext)} \\
  W' &\leftarrow T'[39{:}34] \mathbin\| \widetilde{E}_K^{T}(P) \mathbin\| T'[33{:}0] && \text{(stored word)}
\end{align*}
where $T'$ denotes tweak bits after pipeline registration (unchanged on the 34-bit passthrough portion in practice). One BipBip invocation per 64-bit word suffices; ciphertext does not expand beyond 64 bits.

Figure~\ref{fig:c3-mapping} illustrates the bit layout. Bits $W[33{:}0]$ and the upper six tweak bits $W[63{:}58]$ are not fed through the 24-bit datapath permutation; they provide public context binding the encrypted 24-bit slice to the containing word.

\begin{figure}[ht]
    \centering
    \fbox{\parbox{0.92\linewidth}{\small
    \textbf{Bits 63--58} (6): tweak $T_{\mathrm{hi}}$ \hfill
    \textbf{Bits 57--34} (24): BipBip plaintext $P$ \hfill
    \textbf{Bits 33--0} (34): tweak $T_{\mathrm{lo}}$\\[0.5em]
    \texttt{|}$T_{\mathrm{hi}}$\texttt{|}$P$\texttt{|}$T_{\mathrm{lo}}$\texttt{|} $\leftarrow$ 64-bit cache word $W$;
    stored word replaces $P$ with $\widetilde{E}_K^{T}(P)$, passthrough tweak fields unchanged.}}
    \caption{C$^3$-style decomposition of a 64-bit cache word into 24-bit BipBip plaintext and 40-bit tweak ($T = T_{\mathrm{hi}} \mathbin\| T_{\mathrm{lo}}$). The same layout applies to padded tags on the hit-detection path.}
    \label{fig:c3-mapping}
\end{figure}

\paragraph{Tag path.} The hit detector pads the 52-bit stored tag to 64 bits as $\texttt{cache\_tag} \mathbin\| 0^{12}$, then decrypts with the tag decryptor. The recovered logical tag is $\texttt{decrypted}[63{:}12]$, compared against the incoming address tag delayed by three cycles to align with decryptor latency.

\paragraph{Worked example.} With the fixed verification test key (Appendix~\ref{app:verification}), encrypting \texttt{0x0123456789ABCDEF} yields \texttt{0x0008C70789ABCDEF}: the lower 32 bits are tweak passthrough while the middle slice is permuted (Table~\ref{tab:hwtest}).

\subsection{Pipeline and Controller}
\label{sec:pipelining}

Three BipBip instances serve data encryption (6-cycle latency), data decryption (3-cycle), and tag decryption within hit detection (3-cycle). The 6-cycle encrypt path buffers plaintext while the tweak schedule advances; this is a pipeline scheduling choice, not a requirement of the cipher (Appendix~\ref{app:encryptor}).

\paragraph{Read path.} Tag and data decryptors run in parallel. The incoming address tag is delayed three cycles to align with decrypted tag comparison; plaintext and hit are valid on cycle~3.

\paragraph{Write path.} Plaintext enters the encryptor immediately while the tag path evaluates a hit. Write enable is asserted on cycle~6 only when a hit is confirmed, overlapping the first three encryptor cycles with hit detection (Table~\ref{tab:timing}).

\begin{table}[ht]
\centering
\caption{Cache access timing (cycles after request).}
\label{tab:timing}
\small
\begin{tabular}{lccc}
\toprule
\textbf{Operation} & \textbf{Crypto active} & \textbf{Hit/WE valid} & \textbf{Data valid} \\
\midrule
Read & Decrypt (3) & Cycle 3 & Cycle 3 \\
Write (total) & Encrypt (6) & Cycle 3 (hit) & Cycle 6 (store) \\
Write (effective overhead) & Overlap 0--2 & Cycle 3 & Cycle 6 \\
\bottomrule
\end{tabular}
\end{table}

\section{Related Work}
\label{sec:related}

Table~\ref{tab:cipher-compare} contrasts BipBip with other low-latency primitives used in memory or cache protection. BipBip's small block fits C$^3$-style partial encryption but requires careful mapping for full cache words; wide-block designs (QARMA, PRINCE) simplify line encryption at higher area and latency cost.

\begin{table}[ht]
\centering
\caption{Low-latency ciphers relevant to cache integration (typical published ASIC targets).}
\label{tab:cipher-compare}
\small
\begin{tabular}{lcccc}
\toprule
\textbf{Cipher} & \textbf{Block} & \textbf{Tweak} & \textbf{ASIC latency} & \textbf{Typical use} \\
\midrule
BipBip & 24 & 40 & 3-cycle decrypt & C$^3$ pointers; our cache \\
QARMA-64 & 64 & 64 & $\approx$1--2 cycles & Memory encryption \\
PRINCE & 64 & --- & 1-cycle encrypt & Pervasive crypto \\
MANTIS & 64 & 64 & Low-latency & Memory/tweakable AE \\
SCARF & 10 & --- & Ultra-low & Set randomization \\
\bottomrule
\end{tabular}
\end{table}

Randomized caches (ScatterCache, SCARF) and encrypted caches address orthogonal threats. MATTER~\cite{avanziMATTERWideBlockTweakable2024} and QARMAv2~\cite{avanziQARMAv2FamilyTweakable2023} extend memory-encryption TBCs with stronger bounds. Our prototype does not implement set randomization; future work could combine encrypted lines with SCARF-style index encryption.

\section{Evaluation}
\label{sec:results}

We implemented BipBipCache in VHDL and verified it with AMD Vivado 2025.2 simulation and Xilinx Artix-7 synthesis. Encryptor and decryptor testbenches match the official BipBip C++ reference~\cite{bipbipReferenceCode} on $N{=}5$ vectors (Appendix~\ref{app:verification}); all report \texttt{MATCH}. Pipelined tests confirm 6-cycle encrypt and 3-cycle decrypt latency.

\begin{figure}[ht]
    \centerline{\includegraphics[width=\textwidth]{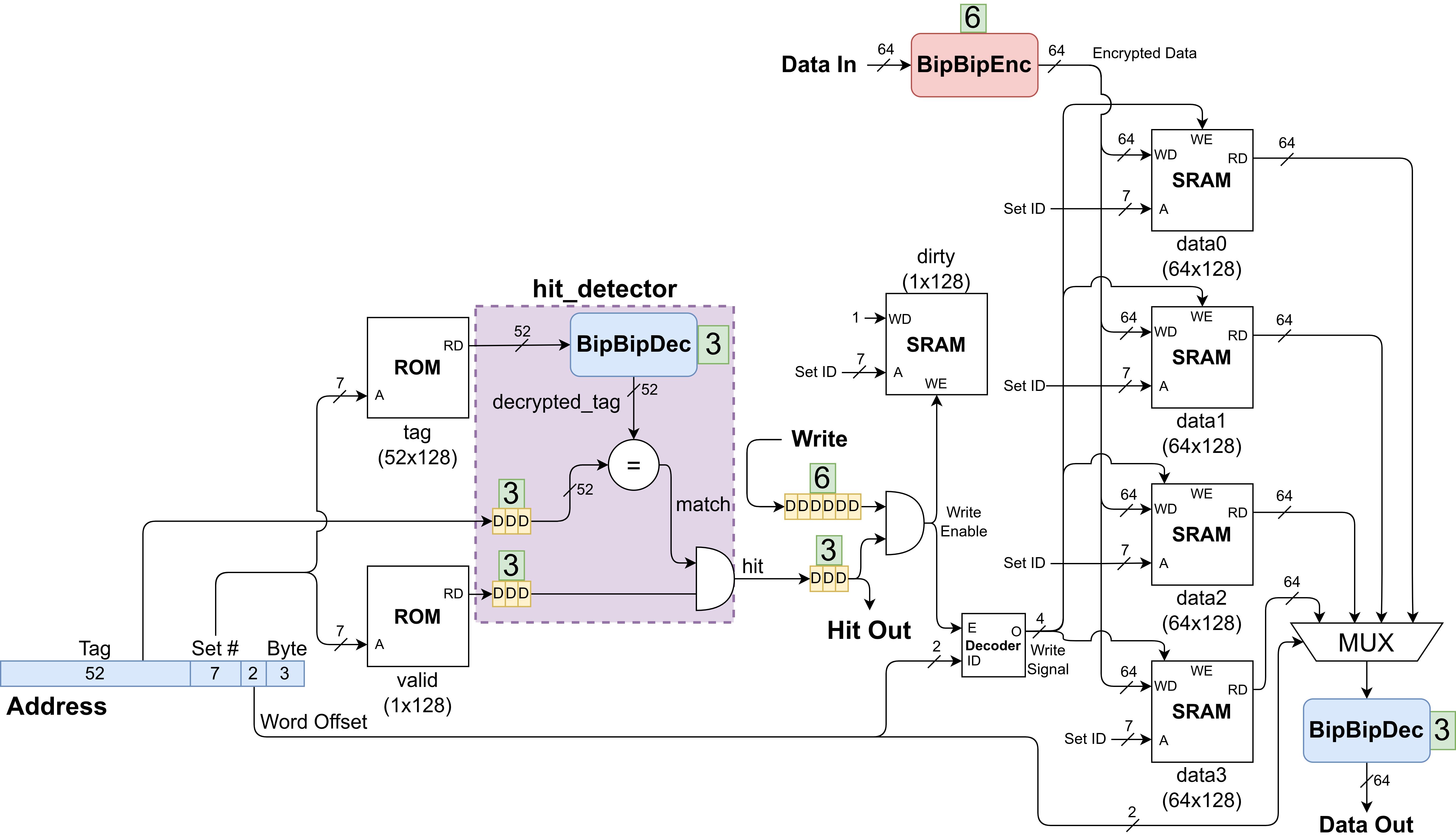}}
    \caption{BipBipCache block diagram: data encryptor, data decryptor, tag decryptor, and direct-mapped SRAM.}
    \label{fig:cache}
\end{figure}

Synthesis targets \texttt{xc7a35tcpg236-1} (Artix-7, 20\,800 LUTs). Figure~\ref{fig:cache} shows the integrated design. Table~\ref{tab:fpga-utilization} summarizes post-implementation resources; cryptographic logic consumes $\approx$79\% of LUTs, indicating that cipher pipelines dominate area on consumer-scale FPGAs.

\begin{table}[ht]
\centering
\caption{Post-implementation resources on Artix-7 (\texttt{xc7a35tcpg236-1}, Vivado 2025.2).}
\label{tab:fpga-utilization}
\small
\begin{tabular}{lrr}
\toprule
\textbf{Resource / module} & \textbf{Used} & \textbf{Utilization} \\
\midrule
Slice LUTs (total) & 3\,356 & 16.1\% of 20\,800 \\
\quad Data encryptor & 1\,185 & --- \\
\quad Data decryptor & 661 & --- \\
\quad Tag path (hit detector) & 818 & --- \\
Slice registers & 1\,500 & 3.6\% of 41\,600 \\
Block RAM tiles & 4 & 8.0\% of 50 \\
\bottomrule
\end{tabular}
\end{table}

The design runs at 100\,MHz in simulation and on a Nexys~A7 board (XC7A100T) with UART-based system tests. Table~\ref{tab:hwtest} summarizes one board-level round-trip.

\begin{table}[ht]
\centering
\caption{Hardware round-trip verification on Nexys~A7.}
\label{tab:hwtest}
\small
\begin{tabular}{lll}
\toprule
\textbf{Stage} & \textbf{Tag} & \textbf{Data} \\
\midrule
Plaintext input & \texttt{0x00000000ABCD1234} & \texttt{0x0123456789ABCDEF} \\
Stored ciphertext & \texttt{0x03FC3D94ABCD1234} & \texttt{0x0008C70789ABCDEF} \\
Decrypted output & \texttt{0x00000000ABCD1234} & \texttt{0x0123456789ABCDEF} \\
\bottomrule
\end{tabular}
\end{table}

\section{Conclusion}
\label{sec:conclusion}

We presented BipBipCache, a feasibility study of embedding BipBip tweakable encryption in an embedded direct-mapped cache. By documenting the C$^3$-style 24+40 bit mapping, reconstructing the first pipelined BipBip encryptor, and overlapping encryption with tag decryption, we show that ultra-low-latency TBCs can be integrated with bounded pipeline overhead. FPGA results on Artix-7 demonstrate practicality for consumer-scale devices, with crypto logic dominating area.

\paragraph{Limitations.} BipBip's theoretical security margin is tight after full-round cryptanalysis~\cite{wangCryptanalysisFullRoundBipBip2024}; only 24 bits per word are strongly permuted; set indices remain cleartext; and we do not evaluate side-channel leakage. Future work includes primitive substitution (e.g., QARMA, MANTIS), set-index protection, formal side-channel analysis, and SoC integration with workload benchmarks.

\appendix

\section{Reference Verification Vectors}
\label{app:verification}

All vectors use a fixed 256-bit test key derived from the ASCII string \texttt{SuperCoolBipBipPasswordForTestin}. Encryptor tests supply ciphertext and expect plaintext; decryptor tests supply plaintext and expect ciphertext.

\begin{table}[ht]
\centering
\caption{Encryptor/decryptor reference vectors ($N{=}5$, all pass).}
\label{tab:verification}
\footnotesize
\begin{tabular}{ll}
\toprule
\textbf{Input} & \textbf{Expected output} \\
\midrule
\texttt{00CD22C000000000} & \texttt{0000000000000000} \\
\texttt{10A93BE49ABCDEF0} & \texttt{123456789ABCDEF0} \\
\texttt{FF7C6AEC76543210} & \texttt{FEDCBA9876543210} \\
\texttt{015D888400000000} & \texttt{0000000400000000} \\
\texttt{FFB9B073FFFFFFFF} & \texttt{FFFFFFFFFFFFFFFF} \\
\bottomrule
\end{tabular}
\end{table}

\section{Encryptor Reconstruction Notes}
\label{app:encryptor}

The BipBip publication specifies decryptor rounds $R$ and $R'$ and tweak functions $G$, $G'$. Our encryptor applies $\mathrm{inverse}(R)$ and $\mathrm{inverse}(R')$ in reverse round order with forward tweak scheduling. Extractors $E_0$, $E_1$ map 53-bit tweak states to 24-bit round keys identically to decryption.

\paragraph{$\theta'$ correction.} An earlier draft claimed that singular $\theta'$ prevented 3-cycle encryption. This was incorrect: on 53-bit states with fixed MSB, $\theta'$ is invertible by back-substitution ($x_{52}=y_{52}$, $x_{51}=y_{51}\oplus x_{52}$, \ldots). The 6-cycle encryptor latency is a pipeline scheduling choice, not a mathematical impossibility.

\bibliographystyle{alpha}
\bibliography{cryptobib/abbrev3,cryptobib/crypto,biblio}

\end{document}